\documentclass[aip,twocolumn,reprint,preprintnumbers,amsmath,amssymb]{revtex4-1}
\usepackage{graphicx}
\usepackage{bm}
\usepackage{dcolumn}
\usepackage{amssymb}
\usepackage[colorlinks,linkcolor=blue,citecolor=blue]{hyperref}
\usepackage{epsfig}
\begin{document}
\title{A ballistic quantum ring Josephson interferometer}
\author{A. Fornieri}
\affiliation{NEST, Istituto Nanoscienze-CNR and Scuola Normale Superiore, I-56127 Pisa, Italy}
\author{M. Amado}
\email{mario.amadomontero@sns.it}
\affiliation{NEST, Istituto Nanoscienze-CNR and Scuola Normale Superiore, I-56127 Pisa, Italy}
\author{F. Carillo}
\affiliation{NEST, Istituto Nanoscienze-CNR and Scuola Normale Superiore, I-56127 Pisa, Italy}
\author{F. Dolcini}
\affiliation{Dipartimento di Scienza Applicata e Tecnologia, Politecnico di Torino, I-10129 Torino, Italy}
\author{G. Biasiol}
\affiliation{CNR-IOM, Laboratorio TASC, Area Science Park, I-34149 Trieste, Italy}
\author{L. Sorba}
\affiliation{NEST, Istituto Nanoscienze-CNR and Scuola Normale Superiore, I-56127 Pisa, Italy}
\author{V. Pellegrini}
\affiliation{NEST, Istituto Nanoscienze-CNR and Scuola Normale Superiore, I-56127 Pisa, Italy}\author{F. Giazotto}
\email{f.giazotto@sns.it}
\affiliation{NEST, Istituto Nanoscienze-CNR and Scuola Normale Superiore, I-56127 Pisa, Italy}
\begin{abstract}

We report the realization of a ballistic Josephson interferometer. The interferometer is made by a quantum ring etched in a nanofabricated two-dimensional electron gas confined in an InAs-based heterostructure laterally contacted to superconducting niobium leads. The Josephson current flowing through the structure shows oscillations with $h/e$ flux periodicity when threading the loop with a perpendicular magnetic field. This periodicity, in sharp contrast with the $h/2e$ one observed in conventional dc superconducting quantum interference devices, confirms the ballistic nature of the device in agreement with theoretical predictions. This system paves the way for the implementation of interferometric Josephson $\pi-$junctions, and for the investigation of Majorana fermions.
\end{abstract}
\maketitle

The capabilities reached by nanoscale fabrication techniques in parallel with the continuous optimization of semiconductor growth and metal deposition approaches have reached a stage at which a wide range of high-performance Josephson-effect-based interferometers can be realized. The relevance of these devices for testing quantum-mechanical phenomena and enabling architectures for quantum computing has been widely recognized.~\cite{Friedman,Hu,Nakamura} Recently, they were also exploited for observing unconventional processes related to heat transport.~\cite{Giazotto-Nature} Future key advances, related to the implementation of interferometric Josephson $\pi-$junctions,~\cite{Dolcini-PRB,Morpurgo} and to the investigation of Majorana fermions~\cite{Alicea,Beenakker-arXiv} would require the realization of Josephson interferometers in which quasiparticles traveling within the normal region of the structure are ballistic.~\cite{Pientka}

To implement ballistic Josephson interferometers two basic requirements must be met: first, the normal/superconductor (NS) contacts should display a resistance sufficiently low that substantial proximity effect takes place in the normal region. Additionally, the electron elastic mean free path in the N-region should be larger that its size. To this end, In$_{x}$Ga$_{1-x}$As (with molar fraction $x\geq 0.75$) semiconductor alloys were widely used in combination with superconducting niobium,~\cite{Schapers-APL,Capotondi} but pure InAs quantum wells were found to be the ideal semiconductor candidates to be coupled to superconductors, thanks to the lack of a Schottky barrier at the contact with a metal, combined with the small effective mass and large spin-orbit coupling.~\cite{Desrat-PRB} Yet, the charge accumulation layer formed at the surface of the InAs~\cite{Smit} was also employed for the realization of low-resistance contacts between the two-dimensional electron gas (2DEG) and a superconducting lead.~\cite{Heida-PRB,Giazotto-Sup} By exploiting nanolithographic techniques, the shape of the N-region comprising a high-mobility 2DEG confined in In$_{x}$Ga$_{1-x}$As-based heterostructure can be tailored at will. Indeed Josephson coupling in superconductor-2DEG-superconductor (S-2DEG-S) junctions was investigated in a variety of configurations like, for instance, quantum point contacts,~\cite{Takayanagi-PRL,Bauch-PRB} quantum dots,~\cite{Deon-APL} or nano-islands.~\cite{Carillo-PRB}

\begin{figure}[t!]
\centerline{\includegraphics[width=\columnwidth,clip=]{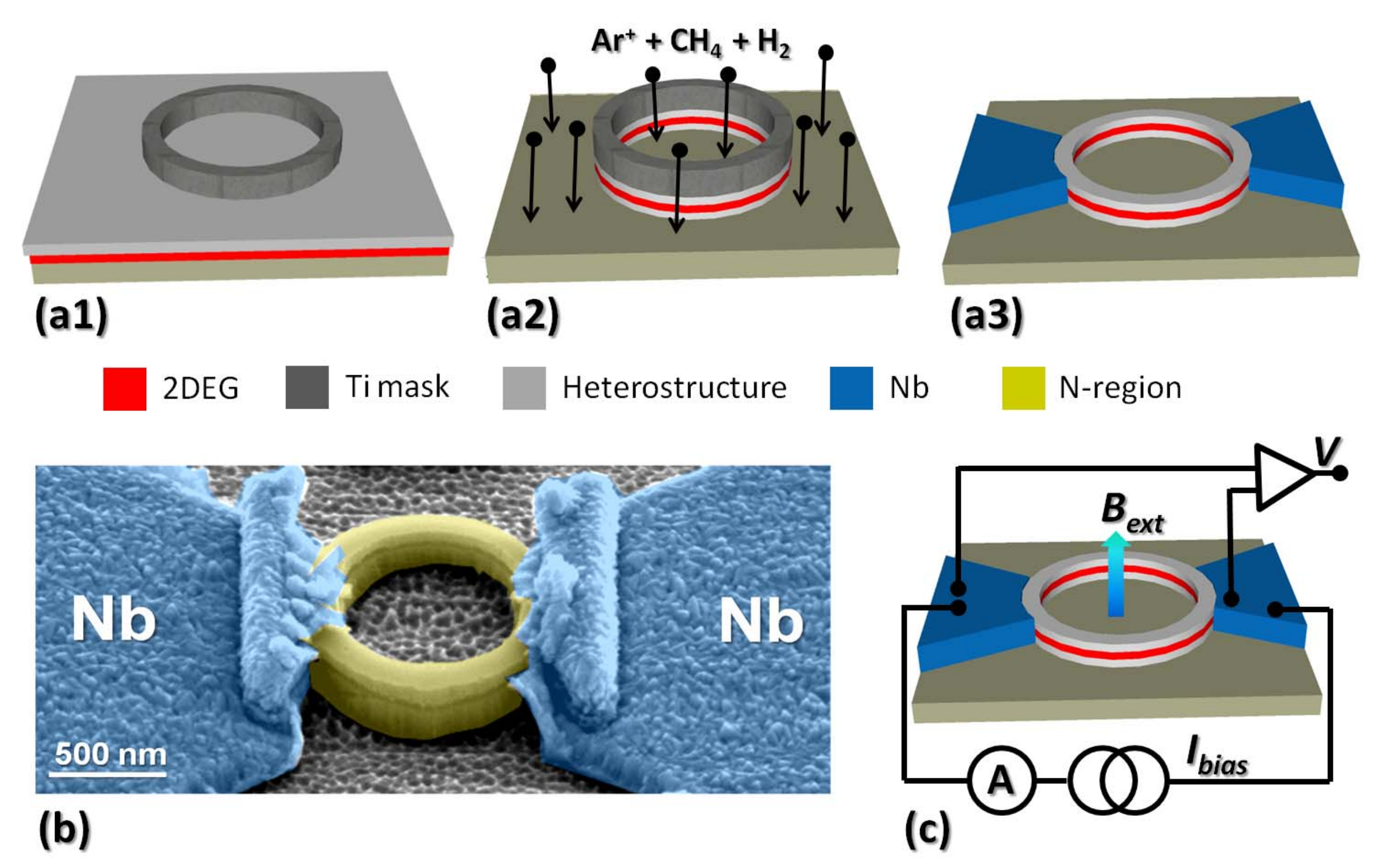}}
\caption{[(a1)-(a3)] Sketch of the nanofabrication steps from the deposition of the Ti-mask (a1), reactive ion etching and definition of the mesa (a2), to the final device consisting of an InAs 2DEG quantum ring (QR) connected to two Nb leads (a3). (b) Pseudo-color scanning electron micrograph of a typical S-QR-S interferometer (scale bar is $500\,$nm). Nb electrodes appear in blue whereas the N-region in dark yellow. (c) Sketch of the 4-wire measurements set-up. $I_{bias}$ is the bias current, $V$ is the voltage drop developed across the junction and $B_{ext}$ is the external magnetic field applied orthogonal to the plane of the ring.}\label{process}
\end{figure}
In this Letter we present the fabrication of a proximized InAs high-mobility 2DEG tailored into a quantum ring (QR) geometry. We demonstrate that our nano-device displays the characteristic fingerprints of a ballistic Josephson interferometer. Indeed, the ballistic N-region of our interferometer  leads to a Josephson coupling characterized by a $h/e$ periodicity typical of the Aharonov-Bohm (AB) effect,~\cite{AB} as theoretically predicted in Ref.~\onlinecite{Dolcini-PRB}, in contrast to the conventional $h/2e$ one observed in analogous metallic dc superconducting quantum interference devices (dc SQUIDs) implemented either with tunnel junctions in parallel,~\cite{Tinkham} or with metallic rings in the diffusive regime.~\cite{Wei-PRB} This device can be sought as a promising building block to implement controllable Josephson $\pi-$junctions,~\cite{Dolcini-PRB,Morpurgo} which are of great interest in quantum computing. In addition, such ballistic superconducting interferometers might pave the way to the experimental investigation of topological superconductors~\cite{Alicea,Beenakker-arXiv,Lutchyn-PRL,Beenakker-NJP} that may support the existence of the elusive Majorana fermions.~\cite{Pientka,Mourik-Sci,Rokhinson-Nat}

Our heterostructure was grown on a (001) GaAs substrate by means of molecular beam epitaxy. A series of $50$-nm-thick In$_{1-x}$Al$_x$As layers were grown on the top of the substrate with the concentration in Al ranging from $x=0.85$ to $x=0.25$. A $4$-nm-thick InAs quantum well was then inserted between two $5.5$-nm-thick In$_{0.75}$Ga$_{0.25}$As layers and In$_{0.75}$Al$_{0.25}$As barriers.~\cite{Capotondi-Thin} The gradient in Al concentration turns out to be essential to avoid lattice mismatch and the resulting strain between the GaAs substrate and the InAs layer.~\cite{Capotondi-Crystal}  The sheet electron density was extracted from Shubnikov-de Haas oscillations at liquid-He temperature without external illumination, and resulted to be $n\simeq6.24\times10^{11}$ cm$^{-2}$. Furthermore, the mobility $\mu\simeq1.6\times10^5$ cm$^2$/V$^{-1}$s$^{-1}$ yields a large elastic mean free path of $l_0\simeq2.3~\mu$m.
\begin{figure}[t!]
\centerline{\includegraphics[width=\columnwidth,clip=]{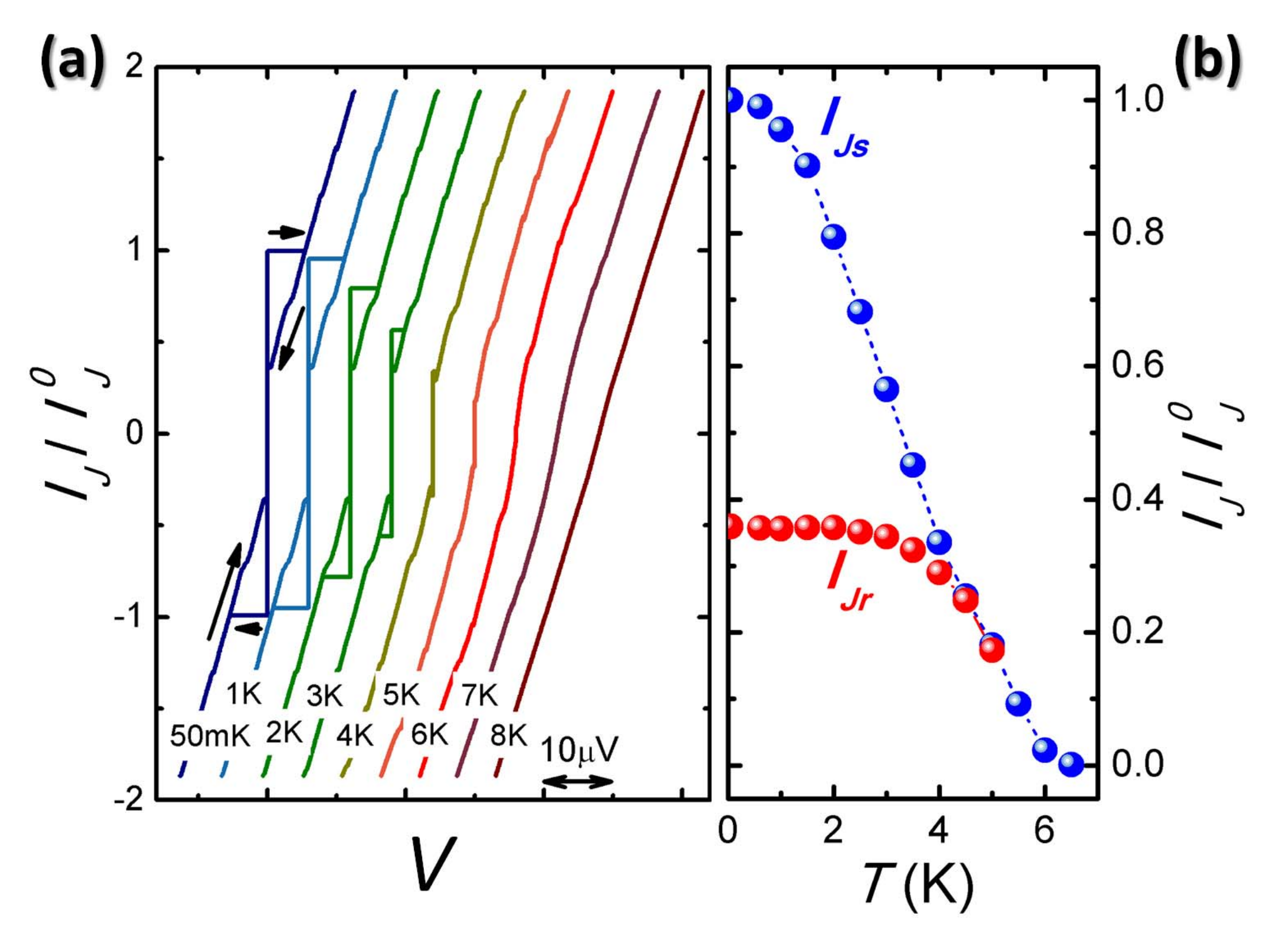}}
\caption{(a) Evolution of the S-QR-S current-voltage ($I-V$) characteristics as a function of temperature $T$ for $B_{ext}=0$. The curves have been horizontally shifted for clarity, and the temperature spans from $50\,$mK to $8\,$K. The Josephson current flowing through the proximized ring survives up to $\sim6.5\,$K, whereas traces of superconductivity in the Nb leads are visible up to $\sim8\,$K. (b) Evolution of $I_{Js}$ and $I_{Jr}$ as a function of $T$. Dotted lines are guides for the eye. $I_{J}^0\simeq52\mu$A is the low-temperature critical current of the junction.}
\label{I_vs_T}
\end{figure}

The nanofabrication of the hybrid S-QR-S Josephson interferometers required a series of three aligned steps of electron beam lithography (EBL), as sketched in Fig.\,\ref{process}[(a1)-(a3)]. Prior the first EBL step the surface of the heterostructure was cleaned in an acetone/isopropanol solution, and baked on a hot plate at $180^\circ$C in order to eliminate adsorbents. Ohmics contacts were obtained in a single EBL operation followed by thermal evaporation of Ni/AuGe/Ni/Au ($10/160/10/60\,$nm), lift-off and annealing at $400^\circ$C for $200\,$s. A subsequent aligned EBL step was done to define the semiconductor mesa into a QR geometry with two lateral protrusions to maximize the contact region between the 2DEG and the superconductors. A $20-$nm-thick Ti mask was deposited by thermal evaporation [see Fig.\,\ref{process}\,(a1)] and the heterostructure was then etched through reactive ion etching in Ar$^{+}$/H$_{2}$/CH$_4$ atmosphere~\cite{Carillo-PE} [see Fig.\,\ref{process}\,(a2)]. The Ti mask was then removed by a $1$:$20$ HF:H$_2$O solution sample rinse. The resulting QR mesa had a half perimeter of $L\simeq1.2~\mu$m ($R\simeq400\,$nm being the mean radius of the QR and its arm width $\sim180\,$nm), smaller than the elastic mean free path $l_0$. The third and last aligned step of lithography [see Fig.\,\ref{process}\,(a3)] was fulfilled to laterally contact the 2DEG with two Nb electrodes, that were deposited by sputtering (with a previous dip into a $1$:$30$ HF:H$_2$O solution and a low-energy Ar$^{+}$ cleaning of the surface) with a growth rate of $\sim1.5\,$nm/s. Our device geometry is technologically simpler with respect to more conventional dc SQUIDs~\cite{Giazotto-NP} because {\it two} different 2DEG-branches share the same {\it two} S-N interfaces in comparison to the {\it four} present in the latter systems. A pseudo-color scanning electron micrograph of one typical device is shown in Fig.\,\ref{process}\,(b).

Low temperature electric measurements were performed in a filtered dilution refrigerator from $50\,$mK up to $8\,$K, where signatures of superconductivity in the Nb leads were still visible. The 4-wire measurement scheme of our experimental set-up is depicted in Fig.\ref{process}\,(c). In particular, the structure is current-biased ($I_{bias}$) whereas the voltage drop ($V$) across the S-QR-S junction is measured with a room-temperature differential preamplifier. Moreover, an external magnetic field ($B_{ext}$) is applied perpendicularly to the 2DEG plane to control the Josephson current interference.

Figure\,\ref{I_vs_T}\,(a) displays the temperature evolution of a few selected S-QR-S current-voltage characteristics at $B_{ext}=0$. In particular, a well defined Josephson current with maximum amplitude of $I_{J}^0\simeq52\mu$A at $50\,$mK is observed, and exhibits a marked hysteretic behavior below $\sim5\,$K. This hysteresis stems from arise from quasiparticle heating in the N-region once the junction switches into the dissipative regime.~\cite{Courtois-PRL} At higher temperature, the hysteresis disappears, so that the switching ($I_{Js}$) and retrapping ($I_{Jr}$) Josephson currents coincide. Traces of superconductivity are visible up to the critical temperature of the Nb electrodes ($T_c\sim8\,$K). The temperature dependence of $I_{Js}$ and $I_{Jr}$ is shown in Fig.\,\ref{I_vs_T}\,(b). Specifically, $I_{Js}$ saturates for $T\lesssim1\,$K whereas $I_{Jr}$ remains constant for a wider range of temperatures till to $T\sim2.5\,$K.

The measured critical temperature yields a value for the superconducting gap of the Nb contacts $\Delta_{Nb}\sim1.764k_BT_c\sim1.2\,$meV ($k_B$ being the Boltzmann constant). In addition, the superconducting coherence length, $\xi_0=hv_F/\Delta_{Nb}\simeq 680\,$nm$\sim (1/2) L$, provides the frame of the {\it intermediate-length} junction ballistic regime. In the above expression, $v_F=9.94\times 10^5$m/s is the Fermi velocity of our InAs 2DEG. The estimate of the excess current,~\cite{Deon-PRB} allows us to deduce the interface scattering parameter~\cite{Flensberg-PRB} $\mathcal{Z}\sim0.4$. The latter leads to a junction normal-state transmissivity~\cite{Blonder} $\mathcal{T}=1/(1+\mathcal{Z}^2)>85\%$, showing the good transparency of the S-2DEG interfaces.
\begin{figure}[t!]
\centerline{\includegraphics[width=\columnwidth,clip=]{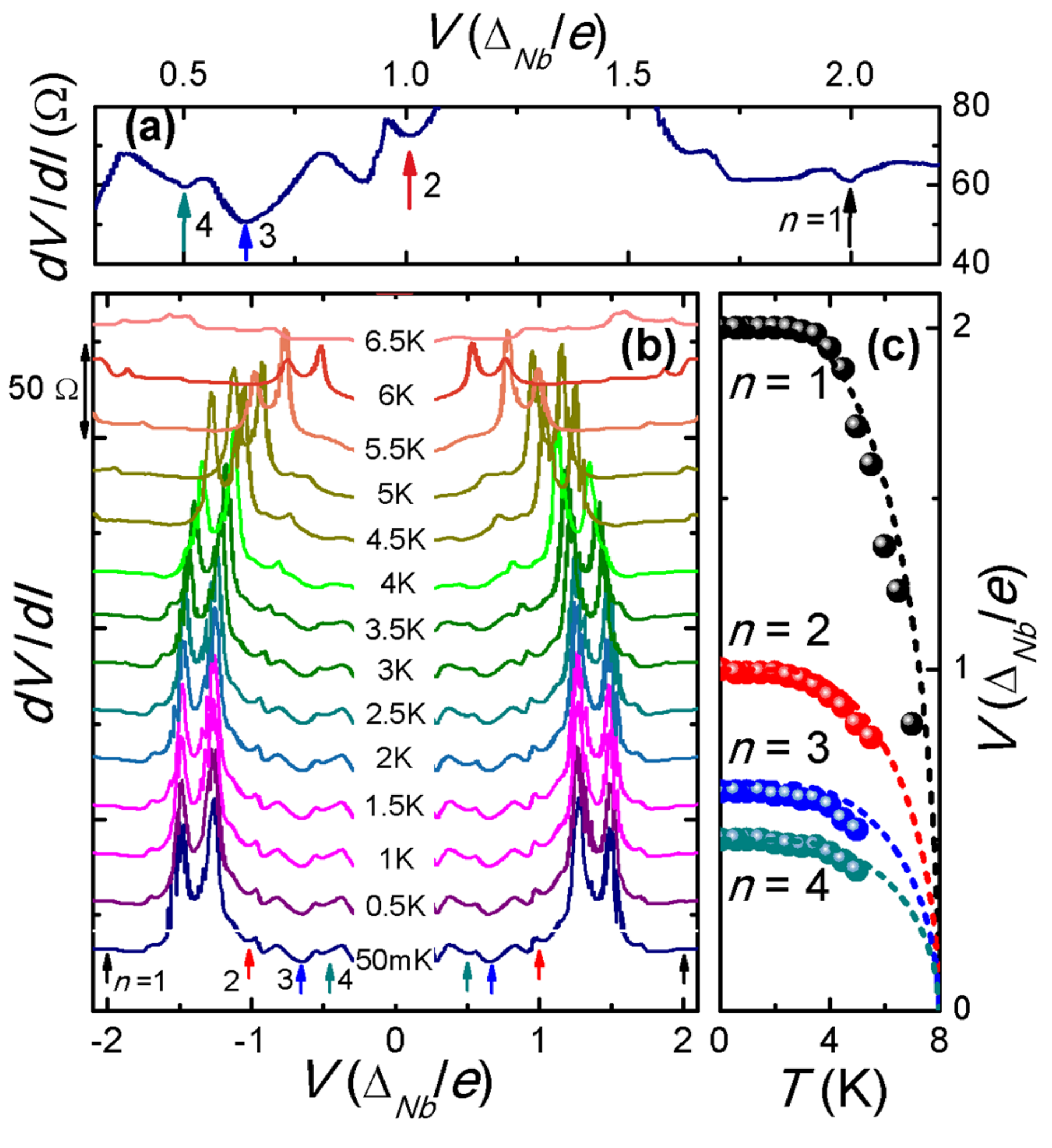}}
\caption{(a) Differential resistance as a function of bias voltage measured at $50\,$mK and $B_{ext}=0$. Colored arrows, placed at $V=2\Delta_{Nb}/en$, match with the first four observable MAR dips. (b) $dV/dI$ as a function of $V$, measured at different temperatures and $B_{ext}=0$. The curves have been vertically offset by $50\,\Omega$ for clarity. (c) Temperature evolution of the position of the first four MAR dips in $dV/dI$. The BCS predictions for the superconducting energy gap are superimposed as dashed lines.}
\label{diff-resistance}
\end{figure}

Figure\,\ref{diff-resistance}\,(a) shows a representative differential resistance ($dV/dI$) trace as a function of $V$ for $B_{ext}=0$ measured at $50\,$mK with low-frequency phase-sensitive lock-in technique. The colored arrows corresponding to voltages $V=2\Delta_{Nb}/en$, where $n$ is an integer, coincide with the position of the first four multiple Andreev reflection (MAR) dips.~\cite{Flensberg-PRB} Figure\,\ref{diff-resistance}\,(b) displays the full temperature evolution of $dV/dI$ versus $V$ at $B_{ext}=0$. In Fig.\,\ref{diff-resistance}\,(c) we show the temperature dependence of the position of the first four MAR dips and the comparison with the Bardeen-Cooper-Schrieffer (BCS) prediction for the superconducting energy gap (dashed lines). We note a good agreement between the BCS prediction and the evolution of the $dV/dI$ dips.
\begin{figure}[t!]
\centerline{\includegraphics[width=\columnwidth,clip=]{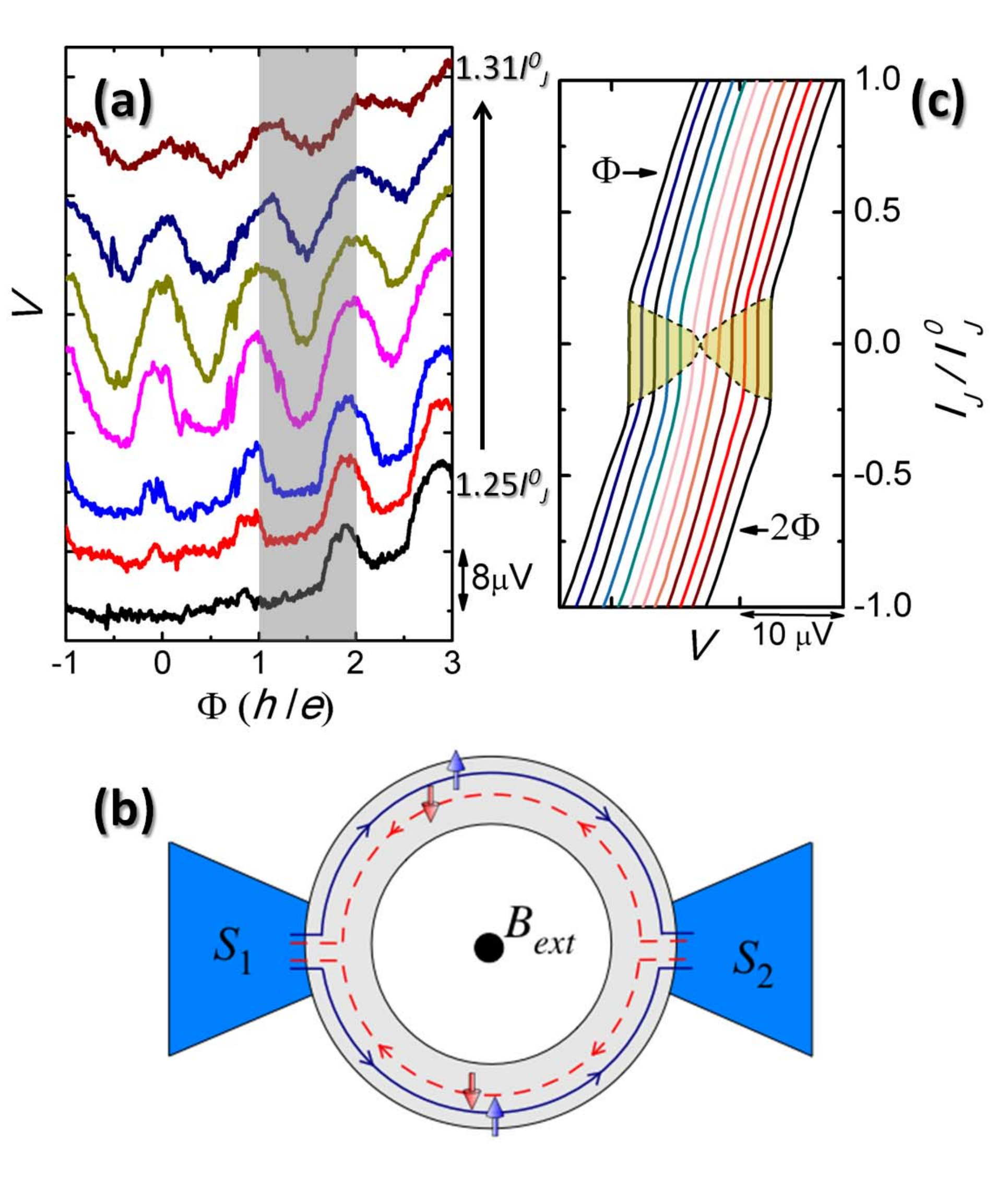}}
\caption{(a) Evolution of the voltage drop developed across the junction vs magnetic flux measured at $50\,$mK for different dc-bias current spanning from $1.25I_J^0$ to $1.31I_J^0$ in steps of $0.01I_J^0$. $V$ shows a modulation with an AB magnetic flux period of $h/e$. The curves have been vertically shifted. (b) Illustration of the two types of Andreev processes occurring in the QR and contributing to the Josephson current interference: an electron with spin $\uparrow$ and a hole with spin $\downarrow$ counter-propagate in the upper (lower) ring arm. The phase difference acquired by traveling in the upper (lower) arm of the ring can be controlled by the magnetic flux and leads to a $h/e$ periodicity in the critical current. (c) Evolution of the $I-V$ characteristics comprised within the hatched gray region in (a). The magnetic flux ranges from $\Phi$ to $2\Phi$ (in units of $h/e$) and increases in steps of $0.083\Phi$. The shadow region comprised within dashed lines highlights the magnetic flux-induced modulation of the Josephson critical current. The curves have been horizontally shifted for clarity and were measured at $5\,$K.}
\label{AB-oscillations}
\end{figure}

We now focus on the measurements that offer evidence that our device performs as a ballistic Josephson interferometer.
To this end we show in Fig.\,\ref{AB-oscillations}\,(a) the evolution of $V$ as a function of $B_{ext}$ measured at $50\,$mK for several dc-bias current values. By increasing $I_{bias}$ the voltage shows an emergent series of peaks whose periodicity matches with the AB period, $\Phi=h/e$, for a QR with $R\simeq306\,$nm. This value for the radius is in good agreement with the inner one of our loop, $\sim310\,$nm. The $h/e$ flux periodicity reveals the ballistic operation of our S-QR-S interferometer as theoretical predicted in Ref.~\onlinecite{Dolcini-PRB}, and differs remarkably from the usual periodicity $h/2e$ present in conventional SQUIDs.

It is worth recalling that the $h/2e$-periodicity originates from the fact that a SQUID is a superconducting ring where {\it two} Josephson weak links having critical currents $I_c^{1,2}$ and phase differences $\chi_{1,2}$, respectively, are set in parallel. The total supercurrent $I_J$ is thus a superposition $I_J=I_c^1 \sin\chi_1 + I_c^2 \sin\chi_2$ of {\it two} Josephson currents that impose a $h/2e$ periodicity. By contrast, the device investigated here consists of {\it one} ring-shaped S-2DEG-S junction. In the simplest case of a single-channel ring in the short-junction limit ($L\ll\xi_0$), the Josephson current is known to be a function of the transmission coefficient $\mathcal{T}$ of the N-region.~\cite{Beenakker-PRL} The resulting $I_J \propto \mathcal{T}  \sin \chi$ (being $\chi$ the phase difference between the two superconductors) oscillates then with the typical AB period $h/e$. More in general (i.e., also away from the short-junction limit), the Josephson current flowing through a SNS junction is carried by Andreev bound states consisting of an electron state correlated to its counterpropagating time-reversed hole state. Because of the ring geometry, each electron (hole) process amplitude has two types of contributions, related to each ring arm, giving rise to four different types of Andreev processes contributing to $I_J$. In particular, the phase difference between the Andreev process where an electron and a hole counter-propagate in the upper arm and the one where an electron and a hole counter-propagate in the lower arm [as depicted in Fig.\,\ref{AB-oscillations}\,(b)] can be controlled by the magnetic flux. This results in a modulation with a period $h/e$ of the critical current, and thereby of the voltage $V$ developed across the junction, confirming our AB periodicity. We also stress that this oscillation period differs from the case of a diffusive S-QR-S junction, where a $h/2e$ periodicity has also been predicted and observed.~\cite{Wei-PRB} We expect that this difference stems from the large number of channels in a diffusive junction which leads $h/2e$ contributions to dominate with respect to the ballistic case.

Figure\,\ref{AB-oscillations}\,(c) shows the junction current-voltage characteristics measured at $5\,$K for several external magnetic flux values comprised within the shadow region in Fig.\,\ref{AB-oscillations}\,(a). The evolution of the Josephson critical current with $\Phi$ is highlighted in light yellow. The range of magnetic fluxes is centered at $1.5\Phi$, thus in the region where the AB voltage oscillations present a minimum. By varying $B_{ext}$, it is possible to tune the Josephson current making it to vanish when approaching $1.5\Phi$ and to reappear when outrunning the minimum. Therefore, this system performs as a superconducting AB interferometer. The curves have been recorded at high temperature, where the supercurrent was small enough to be suppressed by the magnetic flux. At lower temperatures, the Josephson current can be magnetostatically tuned as well [see Fig.\,\ref{AB-oscillations}\,(a)] but not fully suppressed. We stress that our proximized InAs QR Josephson interferometer operates not only in the low-temperature regime but even at temperatures larger than liquid $^4$He, hence widening the number of set-ups that can host this kind of device. In addition, the relatively small magnetic fields required to tune the Josephson coupling makes the implementation of our system even more attractive.

In conclusion, we have demonstrated the fabrication of etched QR-based ballistic Josephson interferometers combining an InAs 2DEG laterally contacted to superconducting Nb leads. The Josephson coupling can be phase controlled by threading the loop with an external magnetic field, and shows a supercurrent periodicity of $h/e$ in sharp contrast with that existing in standard dc SQUIDs. Our S-QR-S system appears as a promising building block for the realization of future controllable Josephson $\pi-$junctions. Further work will be devoted to the implementation of these devices into exotic structures which might favor to the emergence of Majorana fermions.

We wish to thank F. Beltram, F. Deon and F. Taddei for fruitful discussions. Partial financial support from the FP7 program No. 228464 "MICROKELVIN" and from the Italian Ministry of Defence through the PNRM project "TERASUPER" is acknowledged.

\end{document}